\documentclass[aps,pra,superscriptaddress,floatfix,showpacs,10pt,two
column]{revtex4}%
\usepackage[dvips]{graphicx}
\usepackage{amssymb}
\usepackage{amsmath}
\usepackage{color}
\usepackage{amsfonts}%
\definecolor{ueblue}{rgb}{0,0,0.2}
\definecolor{hh}{rgb}{1,0.4,0.3}
\begin{document}
\title{General entanglement-assisted transformation for bipartite pure quantum states}
\author{Wei Song}

\affiliation{Hefei National Laboratory for Physical Sciences at Microscale and Department
of Modern Physics, University of Science and Technology of China, Hefei, Anhui
230026, China}
\author{Yan Huang}
\affiliation{Department of Computer Science and Technology, University of Science and
Technology of China, Hefei, Anhui 230026, China}
\author{Nai-Le Liu}

\affiliation{Hefei National Laboratory for Physical Sciences at Microscale and Department
of Modern Physics, University of Science and Technology of China, Hefei, Anhui
230026, China}
\author{Zeng-Bing Chen}

\affiliation{Hefei National Laboratory for Physical Sciences at Microscale and Department
of Modern Physics, University of Science and Technology of China, Hefei, Anhui
230026, China}
\affiliation{Physikalisches Institut, Universit\"{a}t Heidelberg, Philosophenweg 12,
D-69120 Heidelberg, Germany}


\begin{abstract}
We introduce the \emph{general catalysts} for pure entanglement transformations under local
operations and classical communications in such a way that we disregard the profit and loss of
entanglement of the catalysts \emph{per se}. As such, the possibilities of pure entanglement
transformations are greatly expanded. We also design an efficient algorithm to detect whether a
$k\times k$ general catalyst exists for a given entanglement transformation. This algorithm can as
well be exploited to witness the existence of standard catalysts.

\end{abstract}
\maketitle

Entanglement plays a central role in quantum information processing (QIP)
tasks, such as quantum communication \cite{Bennett:1993}, quantum superdense
coding \cite{Bennett:1992} and quantum computation \cite{Neilsen:2000}. With
the development of quantum information science, people have realized that
quantum entanglement is a kind of physical resource in nature, like energy. To
implement certain QIP tasks, measuring, manipulating and purifying
entanglement \cite{Bennett:1996} by local operations and classical
communications (LOCC) are unavoidable. An important problem concerns the
entanglement transformation between bipartite states under LOCC. This problem
arises as a consequence of the fundamental question of how we can convert one
type of physical resource into another. There have been considerable efforts
devoted to this problem
\cite{Bennett:1997,Nielsen:1999,Vidal:1999,Jonathan:1999,Eisert:2000,Daftuar:2001,Bandyopadhyay:2002,Feng:2005,Duan:2005,Sun:2005}%
. Bennett \textit{et al}. made a first step on this problem
\cite{Bennett:1997} and proposed an entanglement concentration protocol which
solves the entanglement transformation problem in the asymptotic case. Another
significant advance for finite cases was made by Nielsen \cite{Nielsen:1999},
who connected the entanglement transformation with the theory of majorization
\cite{Marshall:1979,Alberti:1982} in mathematics. Let $\left\vert
\psi\right\rangle =\sum_{i=1}^{n}\sqrt{\alpha_{i}}\left\vert i\right\rangle
\left\vert i\right\rangle $ and $\left\vert \varphi\right\rangle =\sum
_{i=1}^{n}\sqrt{\beta_{i}}\left\vert i\right\rangle \left\vert i\right\rangle
$ be pure bipartite states with ordered Schmidt coefficient (OSC) vectors
$\psi=\left(  {\alpha_{1},...,\alpha_{n}}\right)  $ and $\varphi=\left(
{\beta_{1},...,\beta_{n}}\right)  $, where $\alpha_{1}\geq\cdots\geq\alpha
_{n}\geq0$ and $\beta_{1}\geq\cdots\geq\beta_{n}\geq0$. Then there exists a
transformation that converts $\left\vert \psi\right\rangle $ to $\left\vert
\varphi\right\rangle $ with 100{\%} probability under LOCC iff $\psi
\prec\varphi$, where $\prec$ denotes a majorization relation, namely, for
$1\leq l\leq n$
\begin{equation}
\sum\limits_{i=1}^{l}{\alpha_{i}}\leq\sum\limits_{i=1}^{l}{\beta_{i}}.
\label{eq1}%
\end{equation}

Nielsen's theorem provides us with a convenient tool for investigating
entanglement transformation. Shortly after Nielsen's work, a surprising
phenomenon discovered by Jonathan and Plenio \cite{Jonathan:1999} is that
sometimes an extra entangled state can allow otherwise impossible entanglement
transformation to become realizable. The extra state acts just like a catalyst
in a chemical reaction, remaining what it was before the transformation. A
simple example introduced by Jonathan and Plenio is that $\left\vert
\psi\right\rangle \nrightarrow\left\vert \varphi\right\rangle $ but
$\left\vert \psi\right\rangle \otimes\left\vert \chi\right\rangle
\rightarrow\left\vert \varphi\right\rangle \otimes\left\vert \chi\right\rangle
$, where $\left\vert \psi\right\rangle =\left(  {0.4,0.4,0.1,0.1}\right)  $,
$\left\vert \varphi\right\rangle =\left(  {0.5,0.25,0.25,0}\right)  $, and
$\left\vert \chi\right\rangle =\left(  {0.6,0.4}\right)  $. Here, we have used
an OSC vector to represent a bipartite state. But the state {$\left\vert
\chi\right\rangle $} cannot always act as an assistant in the above way, e.g.,
it is not capable of catalyzing the {transformation }$\left\vert
\psi\right\rangle \rightarrow\left\vert \varphi^{\prime}\right\rangle $ where
$\left\vert \varphi^{\prime}\right\rangle =\left(  {0.48,0.27,0.25,0}\right)
$.

However, if we allow some entanglement of the catalyst state $\left\vert
\chi\right\rangle $ to be consumed, then we will greatly improve the
possibilities of entanglement transformations. For example, if we choose
$\left\vert {\chi}^{\prime}\right\rangle =\left(  {2/3,1/3}\right)  $, then,
by using Nielsen's theorem, one can easily verify that the transformation
$\left\vert \psi\right\rangle \otimes\left\vert \chi\right\rangle
\rightarrow\left\vert \varphi^{\prime}\right\rangle \otimes\left\vert {\chi
}^{\prime}\right\rangle $ can be realized under LOCC. During this process,
some entanglement of $\left\vert \chi\right\rangle $ is consumed, i.e.,
$E\left(  {\left\vert \chi\right\rangle }\right)  >E\left(  {\left\vert {\chi
}^{\prime}\right\rangle }\right)  $, where $E$ is the von Neumann entanglement entropy.

Formally, suppose that $\left\vert \psi\right\rangle $ and $\left\vert
\varphi\right\rangle $ are bipartite pure states with $\left\vert
\psi\right\rangle \nrightarrow\left\vert \varphi\right\rangle $ under LOCC,
and that another auxiliary bipartite pure entangled state $\left\vert
\chi\right\rangle $ is standby. If there exists $\left\vert \chi^{\prime
}\right\rangle $ such that $\left\vert \psi\right\rangle \otimes\left\vert
\chi\right\rangle \rightarrow\left\vert \varphi\right\rangle \otimes\left\vert
{\chi}^{\prime}\right\rangle $, we call $\left\vert \chi\right\rangle $ a
\emph{general catalyst} for the entanglement transformation from $\left\vert
\psi\right\rangle $ to $\left\vert \varphi\right\rangle $, in the sense that
we do not care whether the entanglement of catalyst $\left\vert \chi
\right\rangle $ is reduced or increased during the process. When the
entanglement of $\left\vert \chi^{\prime}\right\rangle $ keeps the same as
that of $\left\vert {\chi}\right\rangle $, i.e., $E\left(  {\left\vert
\chi\right\rangle }\right)  =E\left(  {\left\vert {\chi}^{\prime}\right\rangle
}\right)  $, the catalyst $\left\vert \chi\right\rangle $ reduces to the
standard one defined by Jonathan and Plenio \cite{Jonathan:1999}; if $E\left(
{\left\vert \chi\right\rangle }\right)  <E\left(  {\left\vert {\chi}^{\prime
}\right\rangle }\right)  $, the catalyst $\left\vert \chi\right\rangle $
becomes the so called \emph{supercatalyst} introduced by Bandyopadhyay
\textit{et al}. \cite{Bandyopadhyay:2002}; if $E\left(  {\left\vert
\chi\right\rangle }\right)  >E\left(  {\left\vert {\chi}^{\prime}\right\rangle
}\right)  $, some entanglement of the catalyst is consumed, which is
illustrated by the above example; in this case we term $\left\vert {\chi
}\right\rangle $ a \emph{subc}{\emph{atalyst}}.

An important question arises naturally: Given bipartite pure states $\left\vert \psi\right\rangle $
and $\left\vert \varphi\right\rangle $ with $\left\vert \psi\right\rangle \nrightarrow\left\vert
\varphi\right\rangle $ under LOCC, what states can be general catalysts for the above entanglement
transformation? Another question concerns how could we decide whether or not a $k\times k$ general
catalyst exists for certain entanglement transformation. On the other hand, as entanglement is a
very scarce resource (because it cannot be generated by local means and will unavoidably be
degraded by decoherence when transmitted in a noisy environment), we hope that the entanglement of
the catalyst consumed during the entanglement-assisted transformation is as little as possible.
This evokes us to investigate the properties of general catalysts, which will be the first part of
this paper. In the following we shall start with the simplest cases, i.e., entanglement
transformations between $2\times2$ bipartite pure states.

Consider $2\times2$ bipartite pure states $\left\vert \psi\right\rangle ^{\downarrow}=\left(
{\alpha_{1},\alpha_{2}}\right)  $ and $\left\vert \varphi\right\rangle ^{\downarrow}=\left(
{\beta_{1},\beta_{2}}\right)  $, where we have used $\left\vert \psi\right\rangle ^{\downarrow}$ to
denote a state with Schmidt coefficient vectors being sorted in a nonincresing order, and
$\left\vert \psi\right\rangle \nrightarrow\left\vert \varphi\right\rangle $ under LOCC. Assume that
we are provided with another $2\times2$ entangled bipartite pure state $\left\vert
\chi\right\rangle ^{\downarrow}=\left( {x,1-x}\right)  $, where $0.5\leq x\leq1$. The following
theorem provides a \textit{sufficient and necessary} condition for $\left\vert \chi\right\rangle $
to be a general catalyst for realizing the transformation of $\left\vert \psi\right\rangle $ to
$\left\vert \varphi\right\rangle $:

\textit{Theorem 1. The above pure state }$\left\vert \chi\right\rangle
$\textit{ is} a \textit{general catalyst for the transformation of
}$\left\vert \psi\right\rangle $\textit{ to }$\left\vert \varphi\right\rangle
$\textit{ iff }$x\leq{\beta_{1}/\alpha_{1}}$\textit{.}

\textit{Proof}: Assume that there exists $\left\vert {\chi}^{\prime
}\right\rangle ^{\downarrow}=\left(  {{x}^{\prime},1-{x}^{\prime}}\right)  $
such that $\left\vert \psi\right\rangle \otimes\left\vert \chi\right\rangle
\rightarrow\left\vert \varphi\right\rangle \otimes\left\vert {\chi}^{\prime
}\right\rangle $. Using Nielsen's theorem we have $x\alpha_{1}\leq{x}^{\prime
}\beta_{1},$ and so $x\leq x^{\prime}\beta_{1}/\alpha_{1}\leq\beta_{1}%
/\alpha_{1}.$ Conversely, we assume that {$x\leq{\beta_{1}/\alpha_{1}.}$ Then
we shall show that $\left\vert \chi\right\rangle $} is a general catalyst.
Notice first that $\alpha_{1}>\beta_{1},$ because $\left\vert \psi
\right\rangle \nrightarrow\left\vert \varphi\right\rangle .$ For convenience
we divide the problem into two cases.

\textit{Case 1}: $x\geq\alpha_{1}$. In this case we can sort the Schmidt
coefficients of $\left\vert \psi\right\rangle \otimes\left\vert \chi
\right\rangle $ in a nonincreasing order%
\begin{equation}
\left(  {\left\vert \psi\right\rangle \otimes\left\vert \chi\right\rangle
}\right)  ^{\downarrow}=\left(  {\alpha_{1}x,\alpha_{2}x,\alpha_{1}\left(
{1-x}\right)  ,\alpha_{2}\left(  {1-x}\right)  }\right)  .
\end{equation}
If ${x}^{\prime}>\beta_{1}$, then%
\begin{equation}
\left(  {\left\vert \varphi\right\rangle \otimes\left\vert {\chi}^{\prime
}\right\rangle }\right)  ^{\downarrow}=\left(  {\beta_{1}{x}^{\prime}%
,\beta_{2}{x}^{\prime},\beta_{1}\left(  {1-{x}^{\prime}}\right)  ,\beta
_{2}\left(  {1-{x}^{\prime}}\right)  }\right)  ,
\end{equation}
and Nielsen's theorem imposes the following inequalities%

\begin{equation}
\left\{  {%
\begin{array}
[c]{l}%
\alpha_{1}x\leq\beta_{1}{x}^{\prime},\\
x\leq{x}^{\prime},\\
\alpha_{1}+\alpha_{2}x\leq\beta_{1}+\beta_{2}{x}^{\prime}.
\end{array}
}\right.  \label{eq3}%
\end{equation}
These inequalities will be satisfied as long as ${x}^{\prime}\geq\max\left\{
{\frac{\alpha_{1}}{\beta_{1}}x,1-}\frac {\alpha_{2}}{\beta_{2}}(1-x)\right\}  .$ If otherwise,
i.e., ${x}^{\prime}\leq\beta_{1}$, then the second inequality in (\ref{eq3}) should be replaced by
$x\leq\beta_{1}$ which, together with the assumption $x\geq\alpha_{1},$ contradicts with the
requirement $\alpha _{1}>\beta_{1}.$

\textit{Case 2}: $x<\alpha_{1}$. If ${x}^{\prime}>\beta_{1}$, then the
inequalities imposed by Nielsen's theorem are%

\begin{equation}
\left\{  {%
\begin{array}
[c]{l}%
\alpha_{1}x\leq\beta_{1}{x}^{\prime},\\
\alpha_{1}\leq{x}^{\prime},\\
\alpha_{1}+\alpha_{2}x\leq\beta_{1}+\beta_{2}{x}^{\prime},
\end{array}
}\right.  \label{eq4}%
\end{equation}
which hold as long as ${x}^{\prime}\geq\max\left\{  {\frac{\alpha_{1}}{\beta_{1}}x,\alpha_{1}%
,1-}\frac{\alpha_{2}}{\beta_{2}}(1-x)\right\}  .$ If ${x}^{\prime}\leq\beta_{1},$ then the second
inequality in (\ref{eq4}) will be replaced by $\alpha_{1}\leq{\beta}_{1}$ which is a contradiction
to the premise $\alpha_{1}>\beta_{1}.\hfill\blacksquare$

\textit{Remark}. $\left\vert \chi\right\rangle $ is always a subcatalyst since
${x}^{\prime}>x$. If we choose the lower bound of ${x}^{\prime}$, then we
could get an optimal $\left\vert {\chi}^{\prime}\right\rangle $, in that there
will be a minimum loss of entanglement of $\left\vert {\chi}^{\prime
}\right\rangle $. An extreme case on the other side is where the entanglement
of the auxiliary state is completely consumed. Indeed, any $k\times k$
bipartite pure state $\left\vert \chi\right\rangle ^{\downarrow}=\left(
{x_{1},...,x_{k}}\right)  $ is a general catalyst for transforming $\left\vert
\psi\right\rangle $\textit{ to }$\left\vert \varphi\right\rangle $ iff
$x_{1}\leq{\beta_{1}/\alpha_{1}.}$ This can be shown by simply using Nielsen's
theorem and putting $\left\vert \chi^{\prime}\right\rangle ^{\downarrow
}=\left(  {1}\right)  ,$ i.e., a separable state. If $x_{1}$ is strictly
smaller than ${\beta_{1}/\alpha_{1}}$, we can always find sufficiently small
$\varepsilon$ such that $\left\vert \psi\right\rangle \otimes\left\vert
\chi\right\rangle \rightarrow\left\vert \varphi\right\rangle \otimes\left(
{1-\varepsilon,\varepsilon}\right)  $, thereby the auxiliary state will not be
consumed completely.$\hfill\blacksquare$

Next, we consider $3\times3$ cases.

\textit{Theorem 2. Let }$3\times3$\textit{ bipartite pure states }$\left\vert
\psi\right\rangle ^{\downarrow}=\left(  {\alpha_{1},\alpha_{2},\alpha_{3}%
}\right)  $\textit{ and }$\left\vert \varphi\right\rangle ^{\downarrow
}=\left(  {\beta_{1},\beta_{2},\beta_{3}}\right)  $\textit{ be incomparable,
i.e., }$\left\vert \psi\right\rangle \nleftrightarrow\left\vert \varphi
\right\rangle $\textit{ under LOCC. If }$x_{1}\leq\min\left\{  {\beta
_{1}/\alpha_{1},}\left(  {\beta_{1}+\beta_{2}}\right)  /\left(  \alpha
_{1}+\alpha_{2}\right)  \right\}  $\textit{, then an arbitrary }$k\times
k$\textit{ bipartite pure state }$\left\vert \chi\right\rangle ^{\downarrow
}=\left(  {x_{1},...,x_{k}}\right)  $\textit{ is a general catalyst for the
transformation of }$\left\vert \psi\right\rangle $\textit{ to }$\left\vert
\varphi\right\rangle $\textit{.}

\textit{Proof}: Suppose the entanglement of $\left\vert \chi\right\rangle $ is
completely lost after the transformation. Then it suffices to consider the
first two Schmidt coefficients of $\left\vert \psi\right\rangle \otimes
\left\vert \chi\right\rangle $. Two separate cases should be considered in turn.

\textit{Case 1}: $x_{1}\leq x_{2}{\alpha_{1}/\alpha_{2}}$. In this case, the
two largest Schmidt coefficients of $\left\vert \psi\right\rangle
\otimes\left\vert \chi\right\rangle $ are $\alpha_{1}x_{1}$ and $\alpha
_{1}x_{2}$. If there exists $\left\vert \chi\right\rangle $ such that
$\left\vert \psi\right\rangle \otimes\left\vert \chi\right\rangle
\rightarrow\left\vert \varphi\right\rangle \otimes\left\vert \chi^{\prime
}\right\rangle $ with $\left\vert \chi^{\prime}\right\rangle $ being a
separable state, then Nielsen's theorem imposes the conditions that $x_{1}%
\leq{\beta_{1}/\alpha_{1}}$ and $x_{1}+x_{2}\leq\left(  \beta_{1}+\beta
_{2}\right)  /\alpha_{1}$. On the other hand, since $\left\vert \psi
\right\rangle \nleftrightarrow\left\vert \varphi\right\rangle $, it follows
from Nielsen's theorem that one of the following two possibilities must hold:
either
\begin{subequations}
\begin{equation}
\left\{  {%
\begin{array}
[c]{l}%
\alpha_{1}>\beta_{1}\\
\alpha_{1}+\alpha_{2}<\beta_{1}+\beta_{2}%
\end{array}
}\right.  \label{eq6a}%
\end{equation}
or
\begin{equation}
\left\{  {%
\begin{array}
[c]{l}%
\alpha_{1}<\beta_{1}\\
\alpha_{1}+\alpha_{2}>\beta_{1}+\beta_{2}.
\end{array}
}\right.  \label{eq6b}%
\end{equation}
\end{subequations}
In both cases, we have $\beta_{1}+\beta_{2}>\alpha_{1}$. Hence, the condition
$x_{1}+x_{2}\leq\left(  \beta_{1}+\beta_{2}\right)  /\alpha_{1}$ always holds and so can be
neglected.

\textit{Case 2}: $x_{1}>x_{2}{\alpha_{1}/\alpha_{2}}$. By a similar procedure
we can verify that, in order for $\left\vert \chi\right\rangle $ to be a
general catalyst, $x_{1}$ must satisfy $x_{1}\leq\min\left\{  {\frac{\beta_{1}}{\alpha_{1}},\frac{\beta_{1}+\beta_{2}%
}{\alpha_{1}+\alpha_{2}}}\right\}  .$ If the above inequality is strict for all $x_{1}$, then there
exist cases where the entanglement of $\left\vert {\chi}^{\prime}\right\rangle $ is larger than
zero.$\hfill\blacksquare$

The following Theorem 3, Theorem 4, and Example 1 and Example 2 show that sometimes the only
possible choice is to use subcatalysts. This captures what we have emphasized that general
catalysts greatly expand the possibilities of entanglement transformations.

\textit{Theorem 3. Let }$\left\vert \psi\right\rangle $\textit{ and
}$\left\vert \varphi\right\rangle $\textit{ be incomparable states, where
}$\left\vert \psi\right\rangle ^{\downarrow}=\left(  {\alpha_{1}%
,...,\alpha_{n}}\right)  $\textit{ and }$\left\vert \varphi\right\rangle
^{\downarrow}=\left(  {\beta_{1},...,\beta_{n}}\right)  $\textit{. Suppose a
}$2\times2$\textit{ or }$3\times3$\textit{ state }$\left\vert \chi
\right\rangle $ \textit{is a catalyst for the transformation of }$\left\vert
\psi\right\rangle $\textit{ to }$\left\vert \varphi\right\rangle $\textit{. If
}$\alpha_{1}>\beta_{1}$\textit{ and }$\alpha_{n}<\beta_{n}$\textit{, then
}$\left\vert \chi\right\rangle $\textit{ must be a subcatalyst.}

\textit{Proof}: First, suppose $\left\vert \chi\right\rangle ^{\downarrow
}=\left(  {x,1-x}\right)  $ and $\left\vert {\chi}^{\prime}\right\rangle
^{\downarrow}=\left(  {{x}^{\prime},1-{x}^{\prime}}\right)  $. By Nielsen's
theorem we have $\alpha_{1}x\leq\beta_{1}{x}^{\prime}$ which, together with
the condition $\alpha_{1}>\beta_{1}$, implies $x<{x}^{\prime}$. Second,
suppose $\left\vert \chi\right\rangle ^{\downarrow}=\left(  {x_{1},x_{2}%
,x_{3}}\right)  $ and $\left\vert {\chi}^{\prime}\right\rangle ^{\downarrow
}=\left(  {{x}_{1}^{\prime},{x}_{2}^{\prime},{x}_{3}^{\prime}}\right)  $.
Using Nielsen's theorem we have $\alpha_{1}x_{1}\leq\beta_{1}{x}_{1}^{\prime}$
and $\alpha_{n}x_{3}\geq\beta_{n}{x}_{3}^{\prime}$ which, together with
$\alpha_{1}>\beta_{1}$ and $\alpha_{n}<\beta_{n}$, imply $x_{1}<{x}%
_{1}^{\prime}$ and $x_{1}+x_{2}<{x}_{1}^{\prime}+{x}_{2}^{\prime}$.
Consequently, we obtain $\chi\prec{\chi}^{\prime}$.$\hfill\blacksquare$

\textit{Theorem 4. Let }$\left\vert \psi\right\rangle $\textit{ and
}$\left\vert \varphi\right\rangle $\textit{ be }$2\times n$\textit{-level
states with }$\left\vert \psi\right\rangle \nrightarrow\left\vert
\varphi\right\rangle $\textit{, then there does not exist any standard
catalyst or supercatalyst for the transformation of }$\left\vert
\psi\right\rangle $\textit{ to }$\left\vert \varphi\right\rangle $\textit{.}

\textit{Proof}: Suppose there exists a catalyst $\left\vert \chi\right\rangle
$ such that $\left\vert \psi\right\rangle \otimes\left\vert \chi\right\rangle
\rightarrow\left\vert \varphi\right\rangle \otimes\left\vert {\chi}^{\prime
}\right\rangle $, with $E\left(  {\left\vert \chi\right\rangle }\right)  \leq
E\left(  {\left\vert {\chi}^{\prime}\right\rangle }\right)  $. Then we have
$E\left(  {\left\vert \psi\right\rangle }\right)  +E\left(  {\left\vert
\chi\right\rangle }\right)  \geq E\left(  {\left\vert \varphi\right\rangle
}\right)  +E\left(  {\left\vert {\chi}^{\prime}\right\rangle }\right)  $, and
so $E\left(  {\left\vert \psi\right\rangle }\right)  \geq E\left(  {\left\vert
\varphi\right\rangle }\right)  $. Recalling that there is an equivalence
between $\left\vert \psi\right\rangle \rightarrow\left\vert \varphi
\right\rangle $ and $E\left(  {\left\vert \psi\right\rangle }\right)  \geq
E\left(  {\left\vert \varphi\right\rangle }\right)  $ for $2\times
n$\textit{-level states} \cite{Nielsen:1999}, we obtain $\left\vert
\psi\right\rangle \rightarrow\left\vert \varphi\right\rangle ,$ which is a
contradiction.$\hfill\blacksquare$

Note that this theorem is compatible with Theorem 1.

\textit{Example 1}.\textit{ When }$\left\vert \psi\right\rangle $\textit{ has fewer Schmidt
coefficients than} $\left\vert \varphi\right\rangle $, \textit{by using Nielsen's theorem it is
evident that no standard catalyst exists for transforming }$\left\vert \psi\right\rangle $\textit{
to }$\left\vert \varphi\right\rangle $\textit{. However, in some situations the transformation may
be realized by using a subcatalyst. Suppose }$\left\vert \psi\right\rangle ^{\downarrow}=\left(
{\alpha_{1},\alpha_{2}}\right) $\textit{, }$\left\vert \varphi\right\rangle ^{\downarrow}=\left(
{\beta _{1},\beta_{2},\beta_{3}}\right)  $\textit{, and }$\left\vert \chi \right\rangle =\left(
{x,1-x}\right)  $\textit{. To implement the transformation of }$\left\vert \psi\right\rangle
$\textit{ to }$\left\vert \varphi\right\rangle $\textit{, it is obvious that the entanglement of
}$\left\vert \chi\right\rangle $\textit{ should be consumed completely. If
}$x\geq\alpha_{1}$\textit{, then the condition arising from Nielsen's theorem reads
}$x\leq\min\left\{  \beta_{1}/\alpha_{1}{,\beta_{1}+\beta_{2}}\right\} $\textit{; if
}$x<\alpha_{1}$\textit{, then the condition reads }$x\leq \beta_{1}/\alpha,$\textit{
}$\alpha_{1}\leq\beta_{1}+\beta_{2}$\textit{. We
conclude that under the condition }$\alpha_{1}\leq\beta_{1}+\beta_{2}%
$\textit{, the pure state }$\left\vert \chi\right\rangle =\left(
{x,1-x}\right)  $\textit{ with }$x\leq\left\{  \beta_{1}/\alpha_{1}{,\beta
_{1}+\beta_{2}}\right\}  $\textit{ is a subcatalyst for transforming
}$\left\vert \psi\right\rangle $\textit{ to }$\left\vert \varphi\right\rangle
$\textit{.}

\textit{Example 2: Let} $\left| \psi \right\rangle = \left( {\raise0.7ex\hbox{$\mbox{1}$}
\!\mathord{\left/ {\vphantom {\mbox{1}
\mbox{3}}}\right.\kern-\nulldelimiterspace}\!\lower0.7ex\hbox{$\mbox{3}$},\raise0.7ex\hbox{$\mbox{1}$}
\!\mathord{\left/ {\vphantom {\mbox{1}
\mbox{3}}}\right.\kern-\nulldelimiterspace}\!\lower0.7ex\hbox{$\mbox{3}$},\raise0.7ex\hbox{$\mbox{1}$}
\!\mathord{\left/ {\vphantom {\mbox{1}
\mbox{6}}}\right.\kern-\nulldelimiterspace}\!\lower0.7ex\hbox{$\mbox{6}$},\raise0.7ex\hbox{$\mbox{1}$}
\!\mathord{\left/ {\vphantom {\mbox{1}
\mbox{6}}}\right.\kern-\nulldelimiterspace}\!\lower0.7ex\hbox{$\mbox{6}$}} \right)$, \textit{and}
$\left| \varphi \right\rangle = \left( {\raise0.7ex\hbox{$\mbox{1}$} \!\mathord{\left/ {\vphantom
{\mbox{1}
\mbox{6}}}\right.\kern-\nulldelimiterspace}\!\lower0.7ex\hbox{$\mbox{6}$},\raise0.7ex\hbox{$\mbox{1}$}
\!\mathord{\left/ {\vphantom {\mbox{1}
\mbox{6}}}\right.\kern-\nulldelimiterspace}\!\lower0.7ex\hbox{$\mbox{6}$},\raise0.7ex\hbox{$\mbox{1}$}
\!\mathord{\left/ {\vphantom {\mbox{1}
\mbox{6}}}\right.\kern-\nulldelimiterspace}\!\lower0.7ex\hbox{$\mbox{6}$},\raise0.7ex\hbox{$\mbox{1}$}
\!\mathord{\left/ {\vphantom {\mbox{1}
\mbox{6}}}\right.\kern-\nulldelimiterspace}\!\lower0.7ex\hbox{$\mbox{6}$},\raise0.7ex\hbox{$\mbox{1}$}
\!\mathord{\left/ {\vphantom {\mbox{1}
{\mbox{12}}}}\right.\kern-\nulldelimiterspace}\!\lower0.7ex\hbox{${\mbox{12}}$},\raise0.7ex\hbox{$\mbox{1}$}
\!\mathord{\left/ {\vphantom {\mbox{1}
{\mbox{12}}}}\right.\kern-\nulldelimiterspace}\!\lower0.7ex\hbox{${\mbox{12}}$},\raise0.7ex\hbox{$\mbox{1}$}
\!\mathord{\left/ {\vphantom {\mbox{1}
{\mbox{12}}}}\right.\kern-\nulldelimiterspace}\!\lower0.7ex\hbox{${\mbox{12}}$},\raise0.7ex\hbox{$\mbox{1}$}
\!\mathord{\left/ {\vphantom {\mbox{1}
{\mbox{12}}}}\right.\kern-\nulldelimiterspace}\!\lower0.7ex\hbox{${\mbox{12}}$}} \right)$,
\textit{we are provided with another auxiliary }$\mbox{4}\times \mbox{4}$ \textit{entangled
bipartite state }$\left| \chi \right\rangle = \left( {\raise0.7ex\hbox{$\mbox{1}$}
\!\mathord{\left/ {\vphantom {\mbox{1}
\mbox{4}}}\right.\kern-\nulldelimiterspace}\!\lower0.7ex\hbox{$\mbox{4}$},\raise0.7ex\hbox{$\mbox{1}$}
\!\mathord{\left/ {\vphantom {\mbox{1}
\mbox{4}}}\right.\kern-\nulldelimiterspace}\!\lower0.7ex\hbox{$\mbox{4}$},\raise0.7ex\hbox{$\mbox{1}$}
\!\mathord{\left/ {\vphantom {\mbox{1}
\mbox{4}}}\right.\kern-\nulldelimiterspace}\!\lower0.7ex\hbox{$\mbox{4}$},\raise0.7ex\hbox{$\mbox{1}$}
\!\mathord{\left/ {\vphantom {\mbox{1}
\mbox{4}}}\right.\kern-\nulldelimiterspace}\!\lower0.7ex\hbox{$\mbox{4}$}} \right)$ \textit{as
catalyst. We could find an optimal state }$\left| {\chi }' \right\rangle = \left(
{\raise0.7ex\hbox{$\mbox{1}$} \!\mathord{\left/ {\vphantom {\mbox{1}
\mbox{2}}}\right.\kern-\nulldelimiterspace}\!\lower0.7ex\hbox{$\mbox{2}$},\raise0.7ex\hbox{$\mbox{1}$}
\!\mathord{\left/ {\vphantom {\mbox{1}
\mbox{2}}}\right.\kern-\nulldelimiterspace}\!\lower0.7ex\hbox{$\mbox{2}$}} \right)$ \textit{(i.e.,
the entanglement of the subcatalyst state} $\left| \chi \right\rangle $ \textit{consumed during the
transformation reach a minimum value) such that the transformation }$\left| \psi \right\rangle
\otimes \left| \chi \right\rangle \to \left| \varphi \right\rangle \otimes \left| {\chi }'
\right\rangle $ \textit{is possible. Furthermore, it is easy to show that }$\left| \psi
\right\rangle \otimes \left| \chi \right\rangle \leftrightarrow \left| \varphi \right\rangle
\otimes \left| {\chi }' \right\rangle $,\textit{ since the Schmidt coefficients of} $\left| \psi
\right\rangle \otimes \left| \chi \right\rangle $ \textit{and }$\left| \varphi \right\rangle
\otimes \left| {\chi }' \right\rangle $ \textit{are the same. It means that we could also transform
the state} $\left| \varphi \right\rangle \otimes \left| {\chi }' \right\rangle $ \textit{to}
$\left| \psi \right\rangle \otimes \left| \chi \right\rangle $. \textit{Here, we call state}
$\left| \chi \right\rangle $ \textit{as a time-reverse subcatalyst in the above entanglement
transformation process.}

{Next, we consider an interesting question. Let $\left\{  {\left\vert \psi\right\rangle ,\left\vert
\varphi\right\rangle }\right\}  $ and $\left\{ {\left\vert \chi\right\rangle ,\left\vert
{\chi}^{\prime}\right\rangle }\right\}  $ be two incomparable state pairs. Can they assist each
other mutually so as to realize the transformation $\left\vert \psi\right\rangle \otimes\left\vert
\chi\right\rangle \rightarrow\left\vert \varphi\right\rangle \otimes\left\vert
{\chi}^{\prime}\right\rangle $ by LOCC? We shall demonstrate that this can be the case in some
situations.}

\textit{Example 3. Consider two incomparable state pairs }$\left\{ {\left\vert \psi\right\rangle
,\left\vert \varphi\right\rangle }\right\} $\textit{ and }$\left\{  {\left\vert \chi\right\rangle
,\left\vert {\chi }^{\prime}\right\rangle }\right\}  $\textit{, where }$\left\vert
\psi\right\rangle ^{\downarrow}=\left(  {\alpha_{1},\alpha_{2},\alpha_{3}%
}\right)  $\textit{, }$\left\vert \varphi\right\rangle ^{\downarrow}=\left(
{\beta_{1},\beta_{2},\beta_{3}}\right)  $\textit{, }$\left\vert \chi \right\rangle
^{\downarrow}=\left(  {x_{1},x_{2},x_{3}}\right)  ,$\textit{ and
}$\left\vert {\chi}^{\prime}\right\rangle ^{\downarrow}=\left(  {{x}%
_{1}^{\prime},{x}_{2}^{\prime},{x}_{3}^{\prime}}\right)  $\textit{. Suppose that}%
\begin{equation}
\left\{
\begin{array}{l}
 \alpha _2 x_1 \ge \alpha _1 x_2 \ge \alpha _3 x_1 \ge \alpha _2 x_2 ,\alpha
_3 x_2 \ge \alpha _1 x_3 \\
 \beta _2 {x}'_1 \ge \beta _1 {x}'_2 ,\beta _2 {x}'_2 \ge \beta _1 {x}'_3
,\beta _2 {x}'_3 \ge \beta _3 {x}'_1 \\
 \end{array}£¬
 \right.\label{eq11}%
\end{equation}
\textit{Then we can sort the Schmidt coeffcients of }$\left\vert \psi\right\rangle
\otimes\left\vert \chi\right\rangle $\textit{ and }$\left\vert \varphi\right\rangle
\otimes\left\vert {\chi}^{\prime}\right\rangle $\textit{ in a nonincreasing
order }%
\begin{align}
\left(  {\left\vert \psi\right\rangle \otimes\left\vert \chi\right\rangle
}\right)  ^{\downarrow}  &  =\left(  {\alpha_{1}x_{1},\alpha_{2}x_{1}%
,\alpha_{1}x_{2},\alpha_{3}x_{1},\alpha_{2}x_{2},\alpha_{3}x_{2},\alpha
_{1}x_{3},}\right. \nonumber\\
&  \left.  {\alpha_{2}x_{3},\alpha_{3}x_{3}}\right)  ,\\
\left(  {\left\vert \varphi\right\rangle \otimes\left\vert {\chi}^{\prime }\right\rangle }\right)
^{\downarrow}  &  =\left(  {\beta_{1}{x}_{1}^{\prime
},\beta_{2}{x}_{1}^{\prime},\beta_{1}{x}_{2}^{\prime},}\beta_{2}{x}%
_{2}^{\prime},\beta_{1}{x}_{3}^{\prime},\beta_{2}{x}_{3}^{\prime},\beta_{3}%
{x}_{1}^{\prime},\right. \nonumber\\
&  \left.  {\beta_{3}{x}_{2}^{\prime},\beta_{3}{x}_{3}^{\prime}}\right)  .
\end{align}
\textit{Since }$\left\vert \psi\right\rangle \nleftrightarrow\left\vert \varphi\right\rangle
$\textit{, either the set of inequalities in Eq. (\ref{eq6a}) or that in Eq. (\ref{eq6b}) is
satisfied. To be specific, we assume the former. Accordingly, in view of the fact that }$\left\vert
\chi\right\rangle \nleftrightarrow\left\vert \chi^{\prime}\right\rangle ,$\textit{ in order for the
desired entanglement transformation to be
realizable, the following inequalities must be satisfied:}%
\begin{equation}
\left\{
\begin{array}
[c]{l}%
x_{1}+x_{2}>x_{1}^{\prime}+x_{2}^{\prime},\\
\alpha_{1}x_{1}\leq\beta_{1}{x}_{1}^{\prime},\\
\left(  {\alpha_{1}+\alpha_{2}}\right)  x_{1}\leq\left(  {\beta_{1}+\beta_{2}%
}\right)  {x}_{1}^{\prime},\\
\left(  {\alpha_{1}+\alpha_{2}}\right)  x_{1}+\alpha_{1}x_{2}\leq\left(
{\beta_{1}+\beta_{2}}\right)  {x}_{1}^{\prime}+\beta_{1}x_{2}^{\prime},\\
x_{1}+\alpha_{1}x_{2}\leq\left(  {\beta_{1}+\beta_{2}}\right)  \left(
{{x}_{1}^{\prime}+{x}_{2}^{\prime}}\right)  ,\\
x_{1}+\left(  {\alpha_{1}+\alpha_{2}}\right)  x_{2}\leq\beta_{1}+\beta
_{2}\left(  {{x}_{1}^{\prime}+{x}_{2}^{\prime}}\right)  ,\\
x_{1}+x_{2}\leq\beta_{1}+\beta_{2},\\
x_{1}+x_{2}+\alpha_{1}x_{3}\leq\beta_{1}+\beta_{2}+\beta_{3}{x}_{1}^{\prime
},\\
x_{1}+x_{2}+\left(  {\alpha_{1}+\alpha_{2}}\right)  x_{3}\leq\beta_{1}%
+\beta_{2}+\beta_{3}\left(  {{x}_{1}^{\prime}+{x}_{2}^{\prime}}\right)  .
\end{array}
\right.  \label{eq14}%
\end{equation}
\textit{To show these inequalities can be satisfied simultaneously, we choose}%
\begin{align}
\alpha_{1}  &  =0.5,\quad\alpha_{2}=0.26,\quad\alpha_{3}=0.24,\nonumber\\
\beta_{1}  &  =0.49,\quad\beta_{2}=0.48,\quad\beta_{3}=0.03,\nonumber\\
x_{1}  &  =0.62,\quad x_{2}=0.3,\quad x_{3}=0.08.
\end{align}
\textit{Then, the set of inequalities in Eq. (\ref{eq11}) and Eq. (\ref{eq14}) are equivalent to
the
following set:}%
\begin{equation}
\left\{
\begin{array}
[c]{l}%
{x}_{1}^{\prime}\geq31/49,\\
0.97{x}_{1}^{\prime}+0.49x_{2}^{\prime}\geq0.6212,\\
0.97\left(  {{x}_{1}^{\prime}+{x}_{2}^{\prime}}\right)  \geq0.77,\\
0.48{x}_{1}^{\prime}\geq0.49{x}_{2}^{\prime},\\
0.49x_{1}^{\prime}+0.97{x}_{2}^{\prime}\geq0.49,\\
17{x}_{1}^{\prime}+16x_{2}^{\prime}\leq16,\\
{{x}_{1}^{\prime}+{x}_{2}^{\prime}}<0.92.
\end{array}
\right.  \label{sol}%
\end{equation}
\textit{We can picture the region of the independent parameters }%
$x_{1}^{\prime}$\textit{ and }$x_{2}^{\prime}$\textit{ which satisfy the above inequalities
simultaneously in a diagram, Fig. 1. We find the region nonempty,
the desired result. For example, we may choose }$x_{1}^{\prime}=0.81,x_{2}%
^{\prime}=0.1,x_{3}^{\prime}=0.09$\textit{. It should be noted that what }$\left\vert
\psi\right\rangle $\textit{ and }$\left\vert \chi\right\rangle $\textit{ act as in this process are
subcatalysts.}

\begin{figure}[ptb]
\includegraphics[scale=0.65,angle=0]{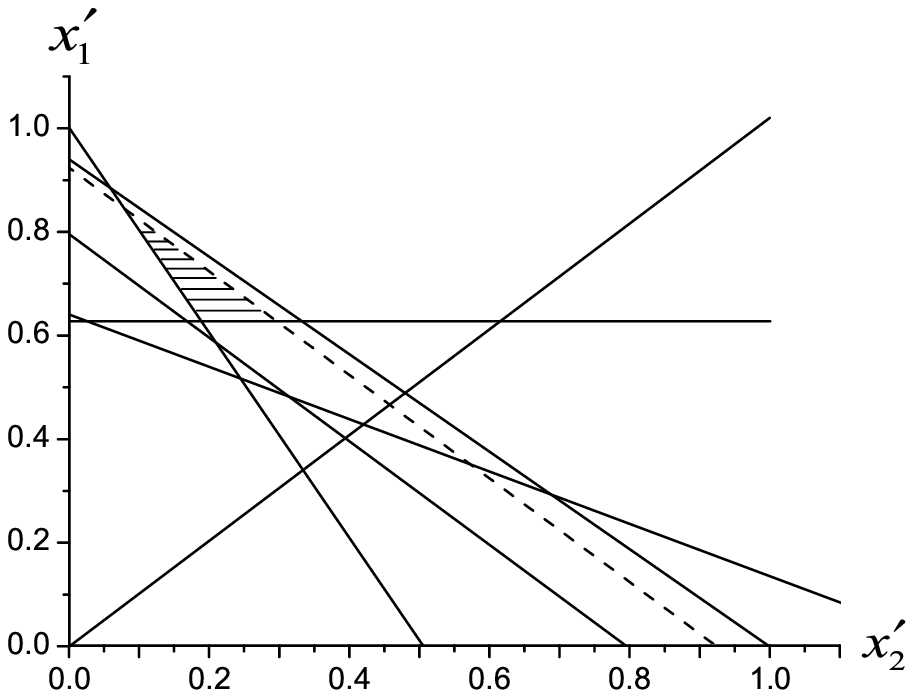}\caption{For two incomparable state pairs $\left\{ {\left| \psi \right\rangle ,\left|
\varphi \right\rangle } \right\}$ and $\left\{ {\left| \chi \right\rangle ,\left| {\chi }'
\right\rangle } \right\}$ with $\left| \psi \right\rangle = \left( {0.5,0.26,0.24} \right)$,
$\left| \varphi \right\rangle = \left( {0.49,0.48,0.03} \right)$, $\left| \chi \right\rangle =
\left( {0.62,0.3,0.08} \right)$, and $\left| {\chi }' \right\rangle = \left( {{x}'_1 ,{x}'_2 ,1 -
{x}'_1 - {x}'_2 } \right)$, if the transformation $\left| \psi \right\rangle \otimes \left| \chi
\right\rangle \to \left| \varphi \right\rangle \otimes \left| {\chi }' \right\rangle $ is feasible,
then the parameters ${x}'_1 $ and ${x}'_2 $ should lie in the shadow region.}
\label{fig1}%
\end{figure}

On the other hand, if $\left\vert \chi\right\rangle \nrightarrow\left\vert
\chi^{\prime}\right\rangle $ and $\left\vert \varphi\right\rangle $ is a
maximally entangled state, then for any state $\left\vert \psi\right\rangle $
we have $\left\vert \psi\right\rangle \otimes\left\vert \chi\right\rangle
\nrightarrow\left\vert \varphi\right\rangle \otimes\left\vert {\chi}^{\prime
}\right\rangle ,$ since otherwise we have\textit{ }$\left\vert \varphi
\right\rangle \otimes\left\vert \chi\right\rangle \rightarrow\left\vert
\psi\right\rangle \otimes\left\vert \chi\right\rangle \rightarrow\left\vert
\varphi\right\rangle \otimes\left\vert {\chi}^{\prime}\right\rangle $. But a
maximally entangled state cannot act as a standard catalyst.

In the remainder of this paper, we will consider the following problem. Assume $\left\vert
\psi\right\rangle ^{\downarrow}=\left(  {\alpha_{1},...,\alpha _{n}}\right)  $ and $\left\vert
\varphi\right\rangle ^{\downarrow}=\left( {\beta_{1},...,\beta_{n}}\right)  $ with $\left\vert
\psi\right\rangle \nrightarrow\left\vert \varphi\right\rangle $ under LOCC. How could we decide
whether or not a $k\times k$ general catalyst exists for converting $\left\vert \psi\right\rangle $
into $\left\vert \varphi\right\rangle $? Notice that it suffices to consider whether the process
$\left| \psi \right\rangle \otimes \left| {\rm X} \right\rangle \to \left| \varphi \right\rangle $
is possible, where $\left| {\rm X} \right\rangle $ is a $k\times k$ maximally entangled state. For
$k \ge n$, the above process is always possible, because a $k\times k$ maximally entangled state
can always be transformed into any $k\times k$ entangled state under LOCC; for $k < n$, we only
need to check whether the majorization relation $\left( {\left| \psi \right\rangle \otimes \left|
{\rm X} \right\rangle } \right)^ \downarrow \prec \left| \varphi \right\rangle ^ \downarrow $
holds. It can be implemented by checking whether the $n - 1$ inequalities are satisfied. However,
this method cannot be applied to decide the existence of $k\times k$ standard catalysts for certain
entanglement transformation. We will propose a Monte Carlo algorithm to solve this problem.
Firstly, we generate a group of $x_1 \ge \cdots \ge x_k \ge 0$ randomly which satisfy
$\sum\nolimits_{i = 1}^k {x_i } = 1$. Then, we merge sort the Schmidt coefficients of $\left| \psi
\right\rangle \otimes \left| \chi \right\rangle $ and $\left| \varphi \right\rangle \otimes \left|
\chi \right\rangle $ in a nonincreasing order where $\left| \chi \right\rangle = \left( {x_1
,\ldots, x_k } \right)$. Now the aim is to check whether the majorization relation $\left( {\left|
\psi \right\rangle \otimes \left| \chi \right\rangle } \right)^ \downarrow \prec \left( {\left|
\varphi \right\rangle \otimes \left| \chi \right\rangle } \right)^ \downarrow $ holds. After
running this procedure $M$ times, if we still cannot find the state $\left\vert \chi\right\rangle $
such that the above majorization relation holds, we say there does not exist a standard catalyst
for this transformation. Of course, there is a failure probability when the algorithm gives a false
output. But when the big number $M$ is large enough, the successful probability of our algorithm
will approach ${1}$. The detailed description is as follows:

(i) For $i=0$ to $k$ do, $x_{i}\leftarrow rand\left[  {0,1}\right]  $, where $x_{1}\geq\cdots\geq
x_{k}\geq 0$, and $\sum_{i=1}^{k}{x_{i}}=1$

(ii) set count$=0$

(iii) while count$<$BIGNUMBER

(iv) begin merge sort the Schmidt coefficients of $\left|  \psi\right\rangle \otimes\left|
\chi\right\rangle $ and $\left|  \varphi\right\rangle \otimes\left|  \chi\right\rangle $ in a
nonincreasing order, respectively

(v) if there exists $\left|  \chi\right\rangle = \left(  {x_{1} ,...,x_{k} } \right)  $ satisfying
$\left(  {\left|  \psi\right\rangle \otimes\left| \chi\right\rangle } \right)
^{\downarrow}\prec\left(  {\left|  \varphi \right\rangle \otimes\left|  \chi\right\rangle } \right)
^{\downarrow}$

(vi) then a $k\times k$ standard catalyst exists for this transformation

(vii) return success

(viii) else

\noindent count=count+1

\noindent end begin

(xi) return failure

In Ref. \cite{Sun:2005} X. M. Sun et al also proposed a deterministic algorithm which runs in
$O\left( {n^{2k+3.5}}\right)  $ time. However, if $k$ is a variable, employing their algorithm to
determine the existence of standard catalyst will become a NP-hard problem. Suppose we choose the
big number to be $M$, then it is easy to see that our algorithm runs in $O\left(  {Mnk}\right)  $
time,which is greatly improved than the deterministic one.

\begin{figure}[ptb]
\includegraphics[scale=0.8,angle=0]{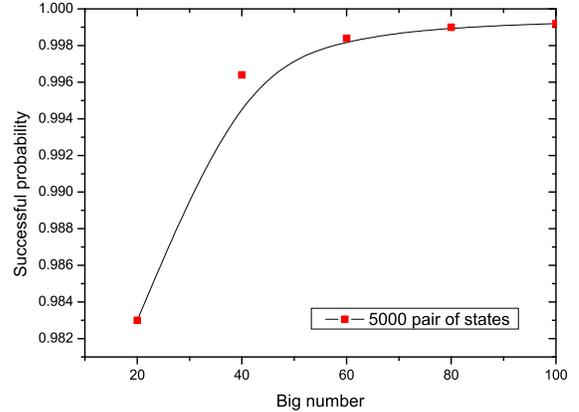}\caption{The numerical results for the successful probability
as a function of the big number we choose. It increases with the big number and  reaches
$99.92{\%}$ when the big number is 100.}
\label{fig2}%
\end{figure}

To show the effectiveness of this algorithm, we will give some examples in the following. We devise
a program to generate $5000$ pair of $8\times8$ states $\left\{  {\left\vert \psi\right\rangle
,\left\vert \varphi\right\rangle }\right\}  $ which always have $4\times4$ standard catalysts,
where $\left\vert \psi\right\rangle \nrightarrow\left\vert \varphi\right\rangle $. We run the above
algorithm and find that, when the big number is chosen to be $100$, the successful probability is
$99.92{\%}$. We plot the result in Fig. 2.

In summary, by introducing the concept of the general catalyst, we can greatly expand the
possibilities of entanglement-assisted transformations between pure entangled states. We consider
the problem of how to decide the existence of a $k\times k$ general catalyst for certain
entanglement transformation. We also propose a Monte Carlo algorithm for determining the existence
of the standard catalyst. When the dimensions of the state and of the potential catalyst are both
very big numbers, our algorithm is far more efficient than the previous deterministic algorithm. We
believe our results may have potential applications in future manipulations of quantum
entanglement.

We thank Shengjun Wu for helpful suggestions. This work was supported by the National NSF of China,
the Fok Ying Tung Education Foundation, and the Chinese Academy of Sciences. We also acknowledge
the support by the European Commission under Contract No. 509487.

\end{document}